\documentclass{iopart}
\usepackage[british]{babel}
\usepackage{alltt}
\usepackage{braket}
\usepackage{graphicx}
\usepackage{float}
\usepackage{ragged2e}
\usepackage[utf8]{inputenc}
\usepackage{lmodern}
\usepackage{microtype}
\usepackage{enumitem}
\usepackage{csquotes}
\usepackage[section]{placeins}
\usepackage{icomma}
\usepackage{booktabs}
\usepackage{nicefrac}
\usepackage{bm}
\usepackage{siunitx}
\usepackage{textcomp}
\usepackage{hyperref}
\usepackage[hyperref=true,
url=false,
isbn=false,
doi=false,
backref=false,
natbib=true,
style=numeric-comp,
sorting=none
]{biblatex}
\bibliography{biblatex.bib}
\AtEveryBibitem{\clearfield{month}}
\AtEveryBibitem{\clearfield{day}}
\usepackage{ulem}
\usepackage{wasysym}

\DeclareSIUnit\pixel{px}

\begin{document}
\title{Spatially resolved spectroscopy of alkali metal vapour diffusing inside hollow-core photonic crystal fibres}
\author{Daniel R Häupl$^{1,2}$, Daniel Weller$^3$, Robert Löw$^3$ and Nicolas Y Joly$^{1,2}$}
\address{$^1$ University of Erlangen-Nürnberg, Staudtstraße 7/B2, 91058 Erlangen, Germany}
\address{$^2$ Max Planck Institute for the Science of Light, Staudtstraße 2, 91058 Erlangen, Germany}
\address{$^3$ 5\textsuperscript{th} Physical Institute, University of Stuttgart, Pfaffenwaldring 57, 70569 Stuttgart, Germany}
\ead{daniel.haeupl@mpl.mpg.de}

\begin{abstract}
We present a new type of compact and all-glass based vapour cell integrating hollow-core photonic crystal fibres. 
The absence of metals, as in a traditional vacuum chamber and the much more compact geometry allows for fast and homogeneous  heating. 
As a consequence we can fill the fibres on much faster timescales, ranging from minutes to hours.
Additionally the all-glass design ensures optical access along the fibre. This allows live monitoring of the diffusion of rubidium atoms inside the hollow-core by measuring the frequency-dependent fluorescence from the atoms. The atomic density is numerically retrieved using a 5-level system of Bloch-equations.
\end{abstract}
\maketitle
\section{Introduction}
Hot atomic vapours are an active field of research and have great potential in the fields of quantum optics~\autocite{caltzidis2021pra, mccormick2007olo} quantum communication~\autocite{finkelstein2018sa, kaczmarek2018pra} or metrology~\autocite{budker2007np, meyer2021pra}.
Optimal atom-light coupling in a vapour cell is limited by the diffraction limit of Gaussian beams. This can be overcome
by confining light and atoms inside a hollow-core fibre.
This led to many applications from nonlinear optics in noble gases~\autocite{travers2011josabj}, to chemistry~\autocite{cubillas2013csr}, and three-photons spectroscopy of caesium vapour inside \textsc{pcf}s~\autocite{epple2014nc}. 
Although previous works~\autocite{ghosh2006prl,light2007olo,epple2014nc} showed the feasibility of filling \textsc{hc-pcf}s with hot vapours, the experimental setup was not optimal. The fibres were placed inside a conventional steel vacuum chamber filled with caesium.
On the one hand, the large size of the steel chamber makes a homogeneous heating of the system not only technically challenging but prone to the creation of cold spots, reducing the vapour density at the position of the fibre. With this it takes accordingly longer until the entire fibre is in equilibrium with the background pressure and homogeneously filled. 
Uneven filling along the fibre length additionally creates severe static electric fields, troublesome for experiments with e.g. Rydberg atoms~\cite{epple2017olo}.
In this paper we present a new type of vapour cell that is entirely made out of glass. 
This includes a new way to mount the fibre inside the cell.
These cells are not much larger than the fibre itself and offer an extended stem. Both regions are encapsuled by an appropriate two-zone oven, preventing condensation in the main part by keeping the stem slightly cooler.
Moreover, since our design allows full optical access from the side, we can monitor the evolution of the atomic density distribution along the whole fibre length through the measurement of radially emitted fluorescence. From spatially resolved fluorescence we can deduce on what timescale the atoms diffuse into the fibre.
Finally we modelled the frequency-dependent absorption of the light propagating inside the hollow-core using a 5-level system to obtain a quantitative understanding of the diffusion process.
At first, we present the concept of the gas-cell and how to hold the fibre properly. 
To study the dynamics of the filling process, we start in an equilibrium situation at a certain temperature and then suddenly raise or drop T and monitor how long it takes to undergo thermalisation.
\section{Experimental setup}
\begin{figure}[ht!]
    \centering
    \includegraphics[width=0.8\linewidth]{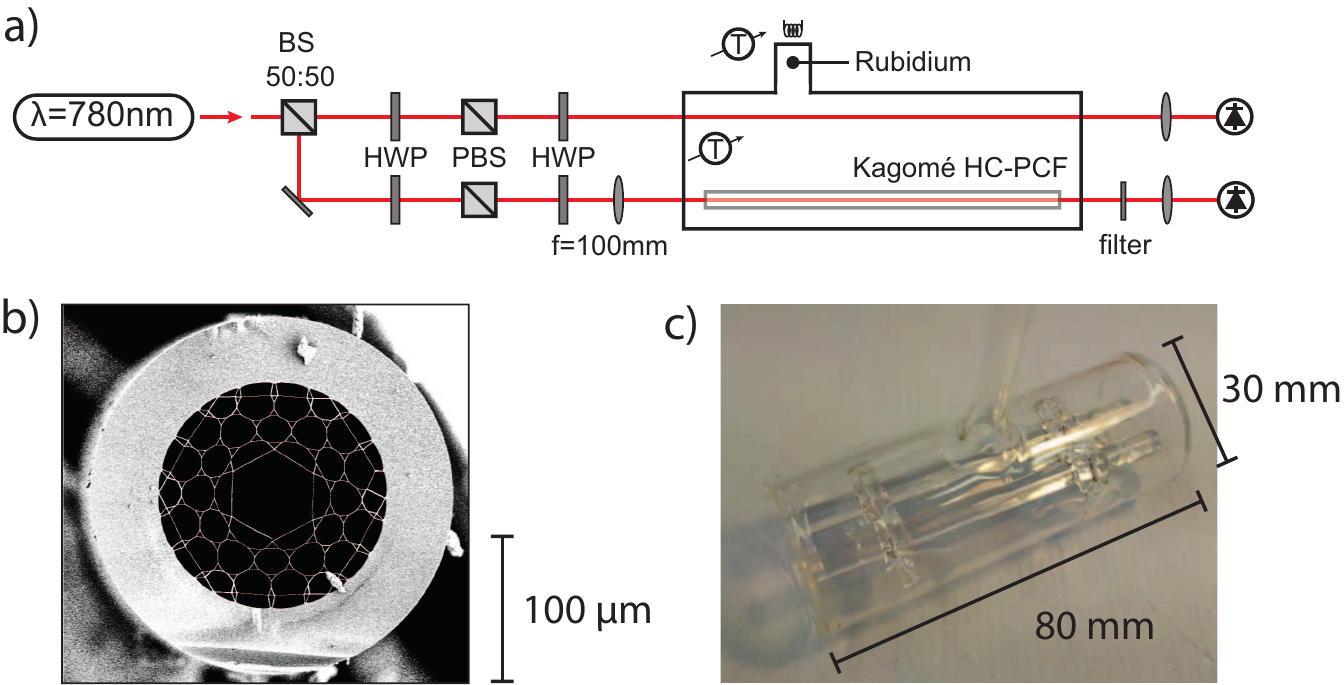}
    \caption{\textbf{(a)} Sketch of the experimental setup. 
    The input beam at \SI{780}{\nm} is initially split equally by a 50:50 beam-splitter BS. 
    Half-wave plates (HWP) and polarisation beam splitter (PBS) ensure the control of the power and a second HWP the input polarisation. 
    One beam is free-space propagating through the cell and the other is coupled into a \SI{60}{\um} core diameter Kagomé hollow-core \textsc{pcf} shown in \textbf{(b)}. 
    At the output of the vapour cell, the beams are separately focused onto photodiodes, with frequency filters in front. 
    \textbf{(c)} is the all-glass vapour cell. 
    The temperature of the reservoir and of the main cell can be addressed independently.}
    \label{fig:setup}
\end{figure}

\begin{figure}[ht!]
    \centering
    \includegraphics[width=0.4\textwidth]{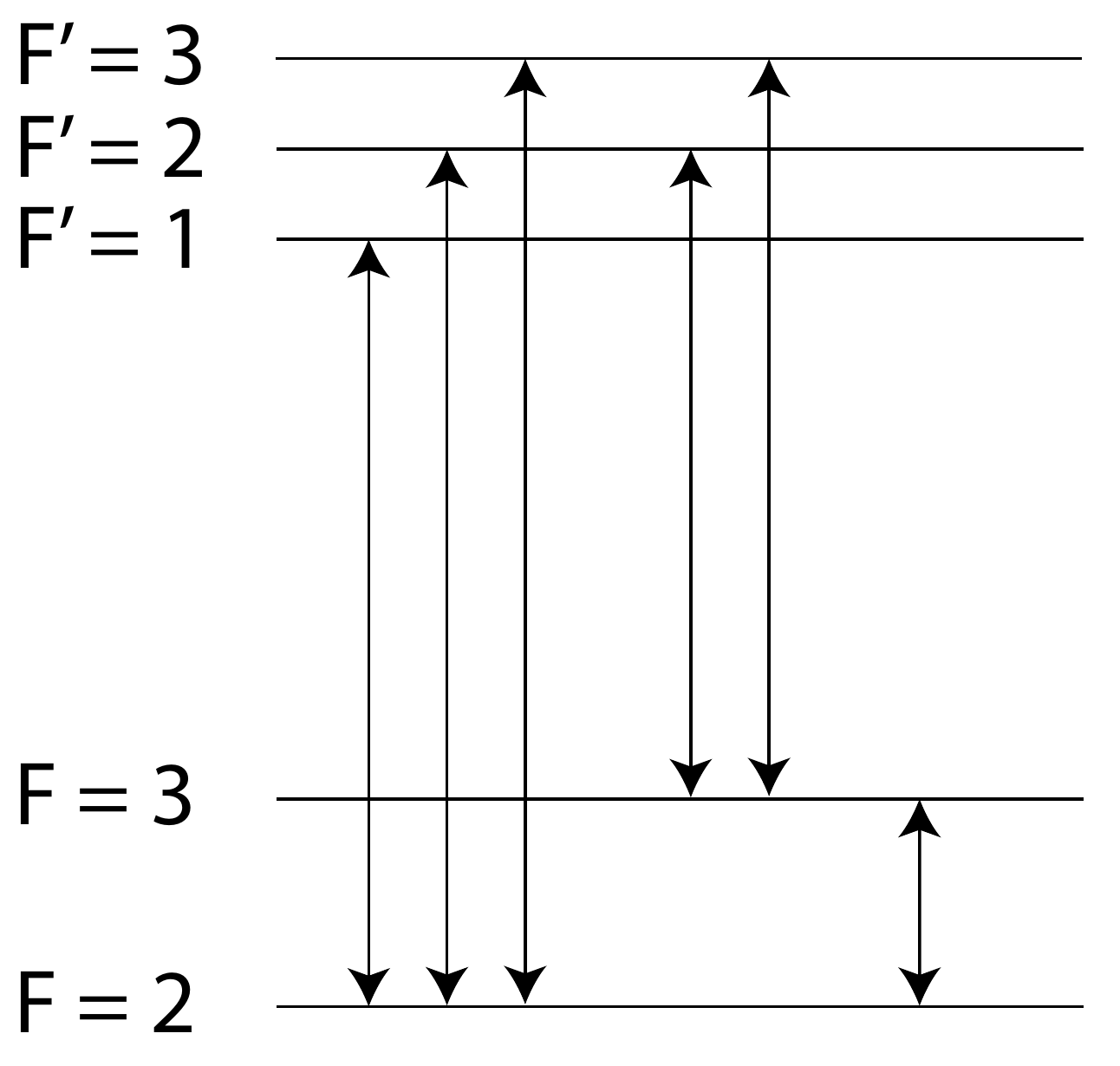}
    \caption{5-level-system used for the retrieval of the atomic density distribution.}
    \label{fig:5_level_system}
\end{figure}

\paragraph{Optical setup.} Our setup is depicted in Fig.~\ref{fig:setup}a. The pump comes from a linearly polarised Toptica DL-pro operated at \SI{780}{\nm}. The pump beam is split into two independent paths, both being sent to the gas-cell. The power of each beam line is controlled by a combination of half waveplate and polarisation beam-splitter cube. We use a second half waveplate to adjust the polarisation state at the entrance of the cell. One beam is coupled to a \SI{60}{\um} core diameter Kagomé fibre using a lens with \SI{100}{\mm} focal distance. 
A scanning electron micrograph of the end face of the fibre is shown in Fig.~\ref{fig:setup}b. 
The fibre is \SI{7.5}{\cm} long. The second beam propagates freely through the gas-cell. 
At the output of the cell we monitor the transmitted power of the two beams with two photodiodes. 
Optical bandpass filters are inserted to prevent ambient light from perturbing the detected signal. 
The glass cell consist of a $\diameter\,\times\,L = 30\,\times\,80\,\text{mm}^2$ tube made of borosilicate where the optical fibre is placed.
To ensure that the fibre is kept straight, we shrunk a \SI{5}{\mm} thick borosilicate tube around the bare optical fibre. 
The fibre-in-tube is not only more sturdy and can be hold in place with two glass supports, but due to the large thickness we are able to distinguish between fluorescence emitted in the hollow-core fibre and the background outside of it. 
This background light stems from leakage out of the fibre, non-perfect coupling and emitted fluorescence. 
Such a design gives access from the side over the complete length of the fibre, allowing side-scattering measurements. 
A secondary $\diameter \,\times\,L = 5\,\times\,45\,\text{mm}^2$ tube is connected at \SI{90}{\degree} to the main cell and acts as the reservoir.
When the cell is sealed before its first use, liquid alkali metal is poured into the reservoir. 
In the current case we filled our cell with rubidium with its natural isotope distribution. 
The temperature of the reservoir and of the main cell can be addressed independently. 
The temperature of the reservoir is kept lower than that of the main cell.
Therefore, not only is the atomic density given by the stem temperature but no Rb atoms will stick to the walls of the main cell.
The diffusion of the rubidium atoms from the reservoir into the main cell is quite fast given the small dimensions and a steady state is achieved almost instantaneously. 
To accelerate the diffusion of the atoms into the fibre, we use elevated temperatures, increasing the overall atomic density.
In principle we could work up to \SI{300}{\degree} or even above, but for technical reason we limit us to the range of 45 to \SI{80}{\degree}.
\paragraph{Filling the vapour cell and the fibre.} 
Ensuring a constant atomic density inside the fibre is of vital interest for experiments involving, e.g. Rydberg states. 
In order to monitor the filling process, we constantly measure the spectra after our fibre and of a free-space beam through the glass cell itself, while scanning the probe wavelength through the complete D2 line of rubidium. 
We target the $^{85}$Rb $\text{F}=2$ line and model it using a Maxwell-Bloch equation approach, including the hyperfine states but neglecting the magnetic sublevels.
In order to take into account the small beam size inside our fibre, we include transit time broadening~\autocite{urvoy2013jpbamop} in our 5-level-system ($^{85}$Rb; $\text{F}=2,3$; $\text{F}^\prime=1,2,3$, \autoref{fig:5_level_system}).
The evolution of the density matrix is governed by the Lindblad master equation~\autocite{steck2007}
\begin{equation}
    i\hbar\frac{\partial \rho}{\partial t}=\left[H, \rho\right]+i\hbar L_D(\rho),
    \label{eq:luiouville_neumann}
\end{equation}
with $H$ the Hamiltonian of the system, $\rho$ the density matrix and $L_D(\rho)$ the Lindblad operator.
The measured spectrum corresponds to the steady state solution of equation \eref{eq:luiouville_neumann}, which yields the excited state fraction of the system.
The off-diagonal elements lead to the complex valued electric susceptibility $\chi$ and thus the absorption coefficient.
To ensure absolute calibration of the frequency-axis, part of the pump beam is guided through a Fabry-Perot interferometer.
For intensities below the saturation intensity, we cross-checked our model to the ElecSus open source software~\autocite{zentile2015cpc} for which length, temperature, magnetic field, light polarization, etc. can be adjusted.
Note that ElecSus does not take power broadening into account, but it is implemented in our model.
In our case, we work above saturation intensity and have to include this in our simulations. 
Beside the intensity, the temperature of the reservoir is the only fitting parameter, which yields the atomic density via the perfect gas law.

\begin{figure}[ht!]
    \centering
    \includegraphics[width=0.8\linewidth]{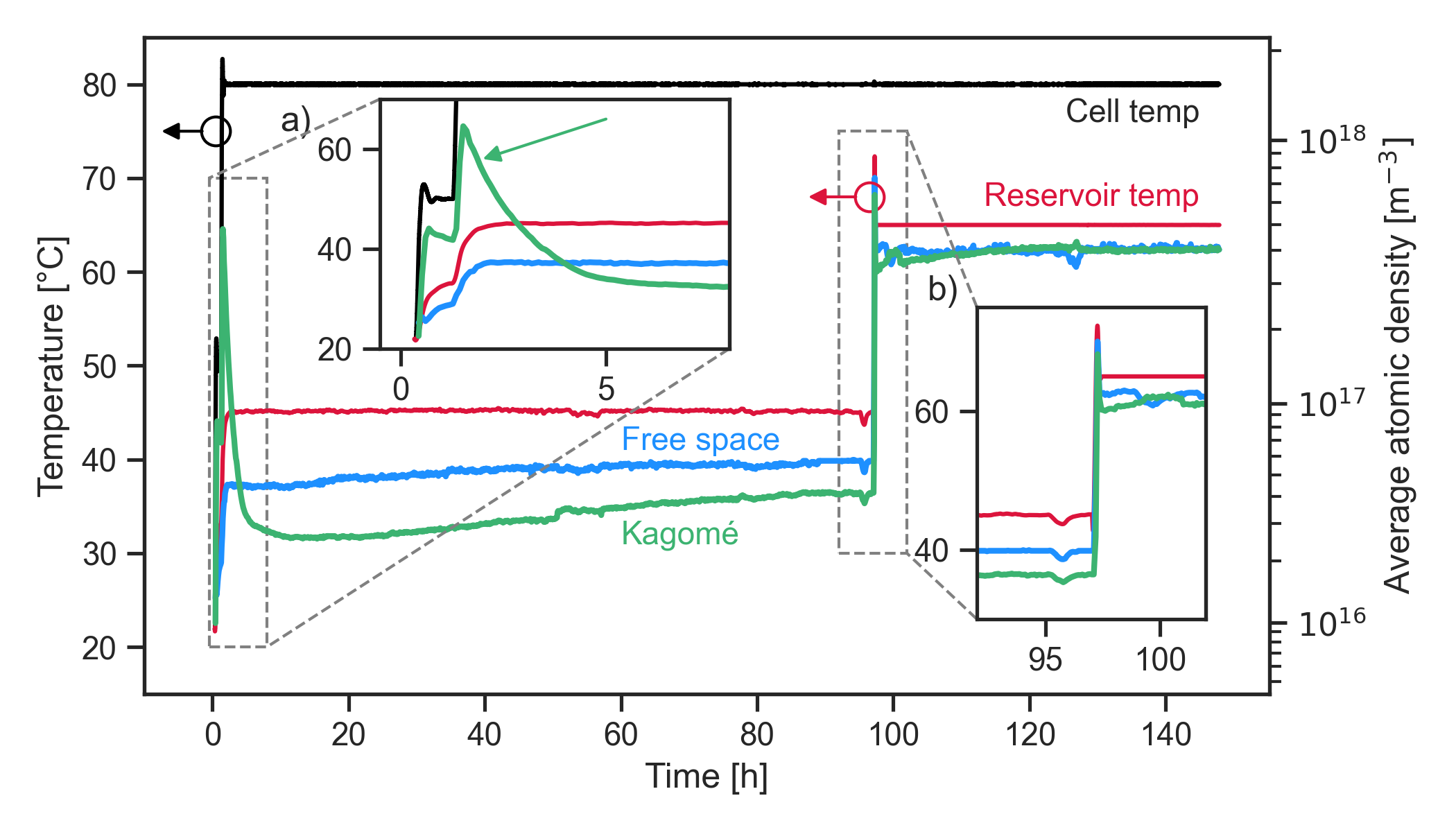}
    \caption{Evolution of the atomic density of rubidium in the vapour cell (blue) and inside the Kagomé fibre (green) vs. the temperature of the cell (black) and of the reservoir (red). 
    The insets magnify the regions where the atomic densities rapidly change due to a modification of the temperatures. The green arrow indicates the apparent overshooting of the density inside the hollow-core when the system is switched on.}
    \label{fig:filling_process}
\end{figure}

\autoref{fig:filling_process} shows the evolution of the atomic density of rubidium in the vapour cell and inside the Kagomé fibre for different temperatures of the cell and of the reservoir. 
At all times we make sure that the temperature of the reservoir remains lower than that of the main cell.
In the initial stage (\autoref{fig:filling_process}a) the vapour cell is heated from room temperature to \SI{80}{\celsius} and left there for \SI{96}{\hour}.
During that time, the reservoir is left at ambient temperature and only heats up due to the conduction of the oven, reaching \SI{45}{\celsius}. 
As shown in \autoref{fig:filling_process}a, as the oven is switched on, the density of atoms in the main cell rapidly increases (blue line).
Surprisingly, the estimation of the density of rubidium retrieved from the spectrum at the output of the fibre overshoots the value that is expected for such a temperature of the reservoir.
We reproducibly observe this whenever we turn on the system.
One possible reason being, that even at room temperature rubidium metal diffused into the fibre and stuck to the core wall.
When we then increase the temperature, those atoms might desorb and after roughly \SI{5}{\hour} the system has relaxed.
Afterwards the density inside the fibre increased with a time constant of \SI{45}{\hour}.
In a second step -- after 96 hours -- we increased the reservoir temperature to \SI{65}{\celsius}, while keeping the cell temperature at \SI{80}{\celsius}.
As seen in \autoref{fig:filling_process}b, the average density in the cell (blue line) follows almost instantaneously the set temperature of the reservoir.
In case of our Kagomé fibre (green line), we also observe this rapid change, but the atomic density reached an equilibrium after $\sim$\,\SI{24}{\hour}, with a timescale of \SI{9.4}{\hour} for the diffusion process.
This should be compared to the previous experiment using a \SI{60}{\um} core diameter Kagomé fibre. 
At that time the filling process lasted $>\!20$\,days~\autocite{epple2017}.

\begin{figure}[h!]
    \centering
    \includegraphics[width=0.8\linewidth]{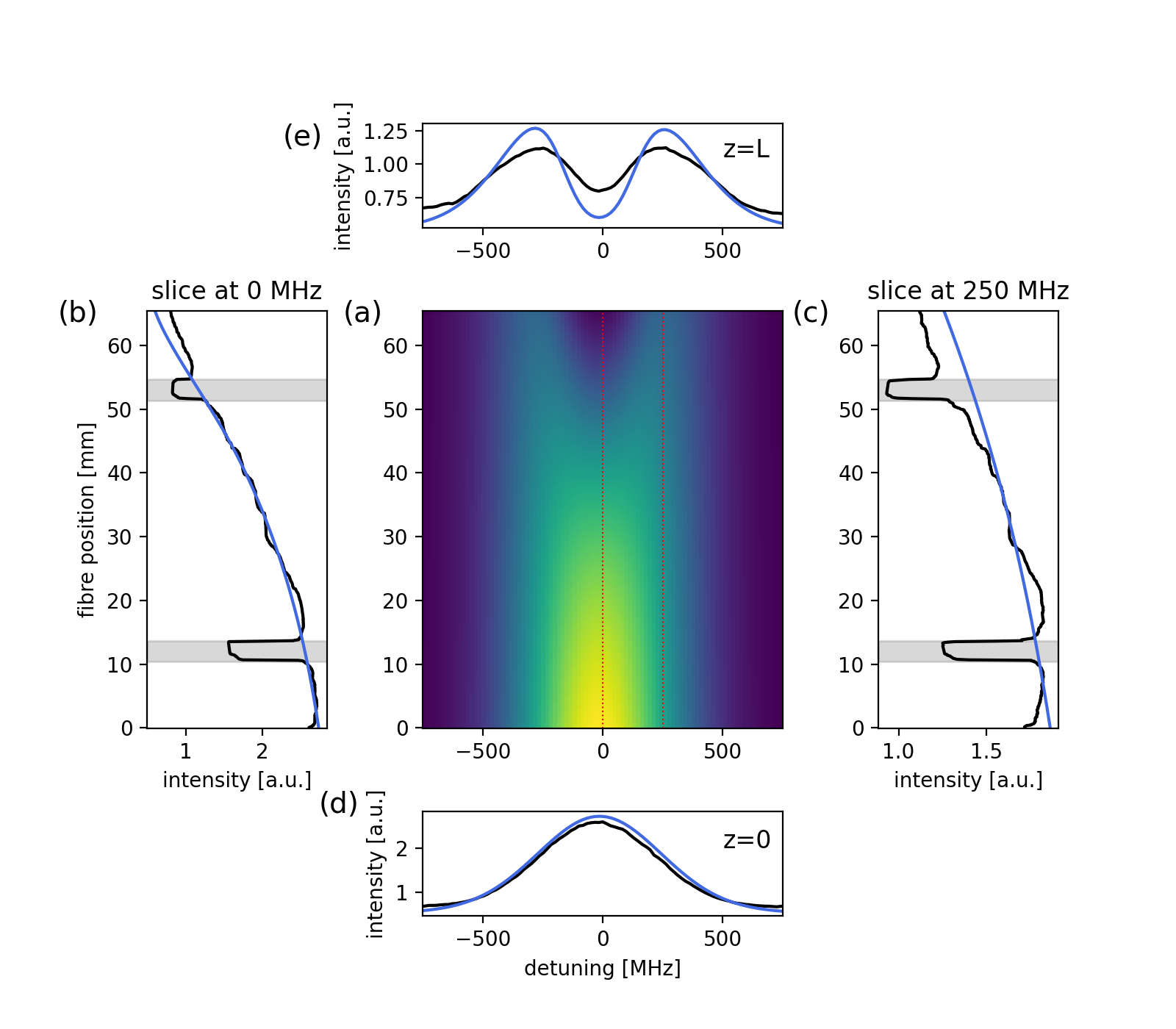}
    \caption{\textbf{(a)} Two-dimensional map of fluorescence emitted by rubidium atoms inside a \textsc{hc-pcf} as a function of detuning and fibre position at \SI{65}{\celsius}.
    Light is entering the cell from the position \enquote{0}. 
    Black indicates experimental data and blue the best fit. 
    The grey area marks the position of the fibre holders, which were ignored for fitting.
    \textbf{(b)} and \textbf{(c)} illustrate the fluorescence along $z$ along at respectively 0 and \SI{250}{\mega\hertz} indicated by the dotted lines.
    \textbf{(d)} and \textbf{(e)} show the spectrum at the beginning ($z=0$) and end ($z=L$) of the fibre.
    }
    \label{fig:2d_plot}
\end{figure}

To even better understand the diffusion in/out of the fibre for an increase/decrease of the gas pressure in the main cell, we additionally look at the radial fluorescence along the fibre.
In a simple picture, the atomic density should be proportional to the emitted fluorescence.
As a proof of principle, we decrease the temperature of the cell, corresponding to a reduction of the atomic density in the surrounding glass cell and causing a density gradient between the fibre and the cell.
We chose a temperature decrease, giving us more time for a careful study of the system dynamics.
Specifically, we decreased the reservoir temperature from 65 to \SI{45}{\celsius} while keeping the cell temperature at \SI{80}{\celsius} and monitored the diffusion process over \SI{100}{\hour}.
While the rubidium diffuses out of the fibre, we take a series of pictures using a 14-bit \textsc{ccd} camera mounted above our glass cell.
To obtain a full spectrum we scan the laser over the the D2 $\text{F}=2$ transition of $^{85}$Rb and image the glass cell every \SI{17.5}{\mega\Hz} of laser detuning.
The input polarisation was chosen so that maximum intensity was recorded by the CCD camera.
The complete scan takes \SI{25}{\second}, which is well below the diffusion timescale that we measured previously (\autoref{fig:filling_process}).
Each image is transversally integrated within the \textsc{roi} to obtain a one-dimensional intensity distribution along the propagation.
By stacking these 160 pictures for all recorded frequencies, we obtain a two-dimensional map of the emitted fluorescence as a function of detuning and fibre position as shown in \autoref{fig:2d_plot}.
The resolution of these maps is around $160\,\times\,870$\,px$^2$ (detuning\,$\times$\,length), depending on the acquisition speed of the camera.
On \autoref{fig:2d_plot}b and c the black lines represent the raw experimental data for two specific frequencies and the blue lines our model.

\begin{figure}[h!]
     \centering
     \includegraphics[width=.8\linewidth]{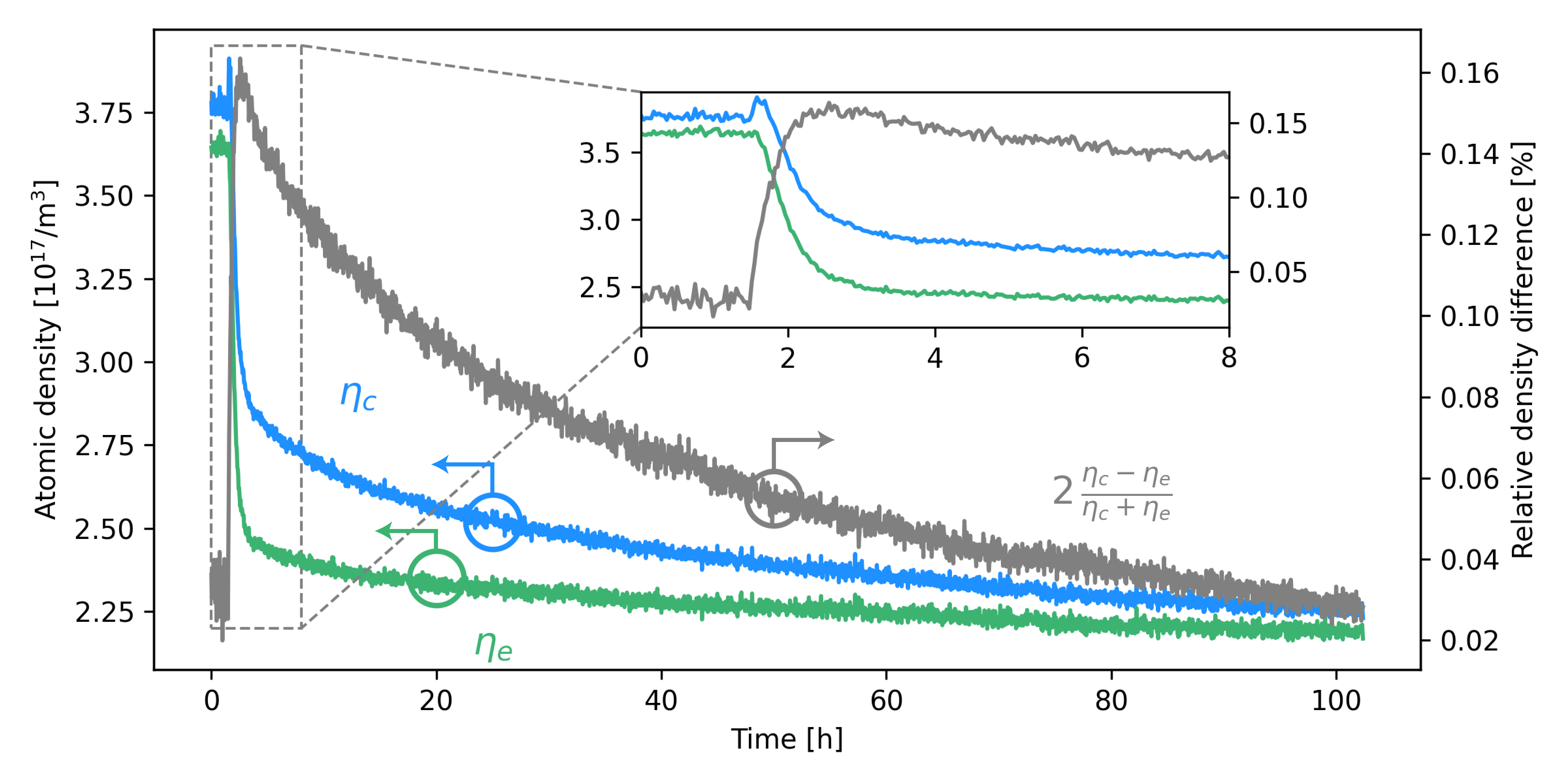}
     \caption{Retrieved atomic densities at the center (blue) and edges (green) of a \SI{60}{\um} Kagomé fibre when decreasing the reservoir temperature from 65 to \SI{45}{\celsius} while keeping a constant cell temperature of \SI{80}{\celsius}.
     The grey curve is the relative density change $2\,\nicefrac{\left(\eta_c-\eta_e\right)}{\left(\eta_c+\eta_e\right)}$.}
	\label{fig:results}
\end{figure}

We extract the evolution of the density inside the fibre for a specific detuning $\Delta$ of the pump using the Beer-Lambert law.
This is done in an iterative way along the whole length of the fibre, assuming an initial intensity $I(z=0)$ and an atomic density $\eta(z=0)$.
As the light propagates the intensity decreases according to
\begin{equation}
    \frac{\mathrm{d}I}{\mathrm{d}z}=\frac{4\pi}{\lambda}
    \,\mathrm{Im}
    \left\{\sqrt{1+\chi\left[\eta(z), I(z), \Delta\right]}\right\}\,I(z)
    \label{eq:beer_lambert}
\end{equation}
where the susceptibility $\chi$ depends on the atomic density $\eta(z)$ and on the light intensity $I(z)$.
As we lower the temperature of the reservoir, the density of atoms inside the cell will decrease, yielding a symmetric diffusion of atoms out of both ends of the fibre.
For simplicity, we assume that the atomic density distribution follows
\begin{equation}
    \eta(z) = (\eta_c - \eta_e) \cdot \left\{1 - \tanh^2\left[\sigma \cdot \left(z - \frac{L}{2}\right)\right]\right\} + \eta_e,
    \label{eq:density_distribution}
\end{equation}
with $L$ the fibre length, $\eta_e$ and $\eta_c$ respectively the atomic density at both ends and at the center of the fibre.
$\sigma$ is chosen so that $\eta(0)=\eta(L)$ is within 1\% of $\eta_e$. This approximately corresponds to a factor of $\sigma=\frac{6}{L}$.
The measured fluorescence (black lines in \autoref{fig:2d_plot}b, c is directly linked with the atomic density and the excited state fraction.
Combining the steady state solution of \eref{eq:luiouville_neumann} with the assumed distribution of \eref{eq:density_distribution}, we calculate $\eta_e$ and $\eta_c$.
As shown on \autoref{fig:2d_plot}b and c the calculated fluorescence curves (blue lines) along the fibre length agree very well with our experimental data. 
\autoref{fig:2d_plot}d and e show the fluorescence spectra at the beginning and the end of the fibre at a given time $t$.
As the diffusion process takes place, we record a series of these maps which are processed the same way.
\autoref{fig:results}a shows the temporal evolution of the atomic density at the end ($\eta_e$) and in the center ($\eta_c$) of the fibre. It is clear that the atomic density at the edge rapidly decreases (green curve) following the change in the cell.
The change in the center (blue curve) is much slower, because the rubidium atoms need to diffuse out of the fibre. 
Like the transmission measurement (\autoref{fig:filling_process}a), the fluorescence measurement shows that the atomic density rapidly changes as soon as the reservoir temperature is modified (\autoref{fig:results}a).
After this abrupt drop the evolution of the atomic density is much slower, with timescales of $\tau_c=\SI{36}{\hour}$ and $\tau_e=\SI{45}{\hour}$, when fitting with a double exponential decay of the form 
\begin{equation}
    \eta(t)=ae^{-t/\tau_1}+be^{-t/\tau_2}+c.
\end{equation}
Here $\tau_c$ respectively $\tau_e$ is the larger of the two lifetimes $\tau_1/\tau_2$, corresponding to the slower diffusion timescale of the system.

To conclude, we presented here an all-glass assembly integrating alkali vapours with a hollow-core photonic crystal fibre.
Its small dimensions and homogeneous temperature permit a rapid modification of the atomic density in a controlled way. 
Additionally, the all-glass design allows to perform easily accessible measurements from the side.
Measuring the fluorescence of the rubidium allows observing in real-time and spatially resolved the diffusion of the atoms out of the fibre when changing the atomic density inside the vapour cell.
Both transmission and fluorescence measurements confirm that the atomic density inside the fibre has reached an equilibrium only after a few days.
This drastically contrasts with previous experiments, where hollow-core fibres were filled with hot alkali metal in bulky metal based vacuum chambers~\autocite{epple2017olo}. 
We believe that such a cell makes an attractive tool for atomic spectroscopy, Rydberg physics and non-linear optics.
Its small size makes this type of cells more transportable and integration in an spectroscopy setup much easier.
In a next step it would be interesting to have an all fibre based setup to avoid free beam coupling to the fibre~\autocite{gutekunst2017ao}.
\section*{Acknowledgements}
This project is supported by the DFG (JO 1090/4-1, LO 1657/6-1 \& RU 1426/1-1) as part of the SPP 1929 ”Giant Interactions in Rydberg Systems (GiRyd)". We also thank SAOT for financial support. 
\printbibliography[heading=bibnumbered]
\end{document}